\documentclass[conference]{IEEEtran}
\IEEEoverridecommandlockouts
\usepackage{cite}
\usepackage{amsmath,amssymb,amsfonts}
\usepackage{algorithmic}
\usepackage{graphicx}
\usepackage{textcomp}
\usepackage{xcolor}
\usepackage{subfigure}
\usepackage{multirow}

\newcommand{\ie}{\textit{i}.\textit{e}.}
\newcommand{\eg}{\textit{e}.\textit{g}.}

\DeclareRobustCommand*{\IEEEauthorrefmark}[1]{%
  \raisebox{0pt}[0pt][0pt]{\textsuperscript{\footnotesize #1}}%
}
\def\BibTeX{{\rm B\kern-.05em{\sc i\kern-.025em b}\kern-.08em
    T\kern-.1667em\lower.7ex\hbox{E}\kern-.125emX}}
\begin{document}

% Efficient Training for Recommendation System through Sparse-To-Dense Paradigm
\title{Alternate Model Growth and Pruning for \\Efficient Training of Recommendation Systems
% \thanks{This work was done during Xiaocong Du’s internship at Facebook. Correspondence to: Xiaocong Du (xiaocong@asu.edu).}
}

\author{
    \IEEEauthorblockN{Xiaocong Du\IEEEauthorrefmark{1}\IEEEauthorrefmark{2}, Bhargav Bhushanam\IEEEauthorrefmark{2}, Jiecao Yu\IEEEauthorrefmark{2}, Dhruv Choudhary\IEEEauthorrefmark{2}, Tianxiang Gao\IEEEauthorrefmark{2}, \\Sherman Wong\IEEEauthorrefmark{2}, Louis Feng\IEEEauthorrefmark{2}, Jongsoo Park\IEEEauthorrefmark{2},  Yu Cao\IEEEauthorrefmark{1}, Arun Kejariwal\IEEEauthorrefmark{2}}
    \IEEEauthorblockA{\IEEEauthorrefmark{1} Arizona State University, Tempe, AZ, USA}
    \IEEEauthorblockA{\IEEEauthorrefmark{2} Facebook, Inc. Menlo Park, CA, USA}
    Email: \{xiaocong, ycao\}@asu.edu, \\ \{bbhushanam, jiecaoyu, choudharydhruv, clingsz, shermanwong, lofe, jongsoo, akejariwal\}@fb.com
}

% City, Country, email address or ORCID}
% \and
% City, Country, email address or ORCID}
% \and
% City, Country, email address or ORCID}
% \and
% \IEEEauthorblockN{4\textsuperscript{nd} Given Name Surname}
% City, Country,email address or ORCID}
% \and
% \IEEEauthorblockN{5\textsuperscript{rd} Given Name Surname}
% City, Country, email address or ORCID}
% \and
% \IEEEauthorblockN{6\textsuperscript{rd} Given Name Surname}
% City, Country, email address or ORCID}
% }

\maketitle

\begin{abstract}
Deep learning recommendation systems at scale have provided remarkable gains through increasing model capacity (\ie~wider and deeper neural networks), but it comes at significant training cost and infrastructure cost. Model pruning is an effective technique to reduce computation overhead for deep neural networks by removing redundant parameters. However, modern recommendation systems are still thirsty for model capacity due to the demand for handling big data. Thus, pruning a recommendation model at scale results in a smaller model capacity and consequently lower accuracy. To reduce computation cost without sacrificing model capacity, we propose a dynamic training scheme, namely alternate model growth and pruning, to alternatively construct and prune weights in the course of training. Our method leverages structured sparsification to reduce computational cost without hurting the  model capacity at the end of offline training so that a full-size model is available in the recurring training stage to learn new data in real time.  To the best of our knowledge, this is the first work to provide in-depth experiments and discussion of applying structural dynamics to recommendation systems at scale to reduce training cost. The proposed method is validated with an open-source deep learning recommendation model (DLRM) and the state-of-the-art  industrial-scale production models.

\end{abstract}

\begin{IEEEkeywords}
Recommendation system, model growth, model pruning, deep learning, deep neural network, efficient model training

\end{IEEEkeywords}

\section{Introduction} \label{sec:intro}
% 3 columns
In the past decade, deep learning recommendation models have successfully powered various applications and products. For instance, YouTube adopts deep candidate generation and ranking models to deliver video suggestions~\cite{covington2016deep}; Microsoft utilizes deep learning recommendation models for news feed~\cite{elkahky2015multi}; Alibaba leverages recommendation models for product suggestions~\cite{zhou2019deep}; Facebook uses deep learning recommendation models to support personalization and ranking in widespread products~\cite{gupta2020deeprecsys,gupta2020architectural,instagram}. 
Recent years, along with the emergence of big data, recommendation models are rapidly expanding their capacity, especially the industry recommendation systems. For example, the capacity for training recommendation systems at web-scale internet companies quadruples~\cite{naumov2020deep} over the past two years, and more than half of the AI training cycles is devoted to training deep learning recommendation models.

%%% ================================================================================
\begin{figure}[!t]
    \centering
    \includegraphics[width=1.0\columnwidth]{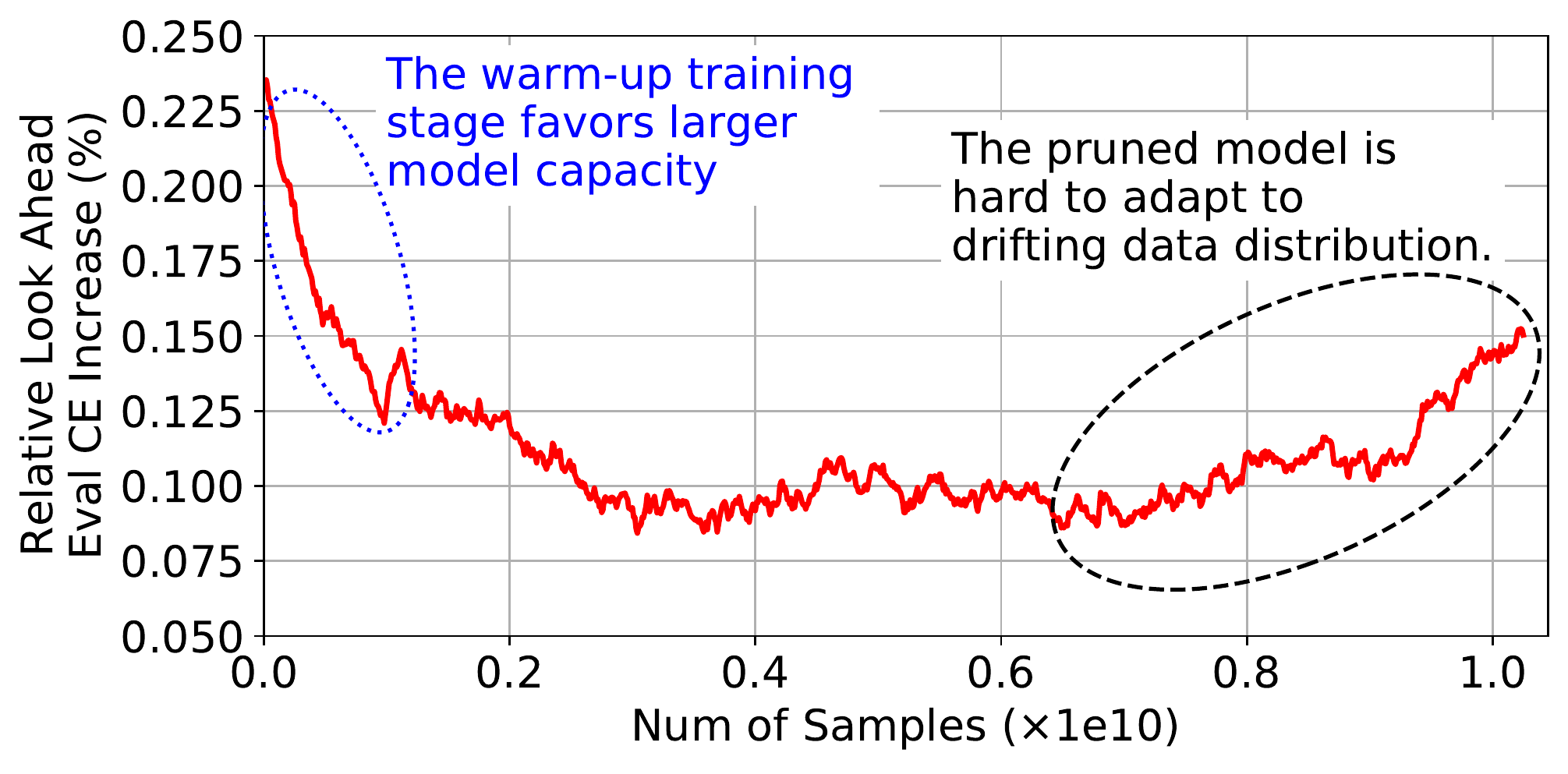}
    \caption{Relative window cross entropy loss  increase (\%) for a pruned recommendation model with 80\% sparsity compared to a full-size model. Unlike other domains, pruning is not the optimal compression technique for an industrial-scale recommendation model.}
    \label{fig:pruning}
\end{figure}
%%% ================================================================================

Training these recommendation systems at scale requires increasingly massive computation power and energy consumption~\cite{naumov2020deep}.  Similar to modern deep neural networks (DNNs) in general domains (such as computer vision)~\cite{han2016dsd,han2015learning,gupta2020compression}, significant computation expense in recommendation area is chiefly due to the model scaling-up, \ie~wider and deeper network with a large number of parameters. Previously, compression techniques, such as pruning~\cite{han2015learning}, quantization~\cite{choi2018pact}, and distillation~\cite{polino2018model}, have been developed~\cite{gupta2020compression} to relieve general DNNs from expensive computational cost caused by over-parameterization. Among them, pruning has been validated as a simple yet efficient method to reduce the redundancy of general DNNs. The pruning technique removes less important parameters from DNN after a complete training on the input data. It then fine-tunes the sparse model with some fine-tuning data, generating a final model with sparsity (\ie~decent amounts of parameters are set to zero). The success of pruning relies on the following priors: (1) the DNN itself is over-parameterized, in other words, the number of parameters is redundant for the given input data~\cite{gordon_2019}; (2) the fine-tuning stage contains large amounts of fine-tuning data to recover the accuracy drop caused by pruning; (3) the training data and fine-tuning data follow the same distribution; (4) the pruned model is usually served for the purpose of inference, which has less demand on model capacity since learning new data is not required.

Unlike general DNNs, industrial-scale deep learning recommendation systems have the following properties that are unique:
    \paragraph{Insufficient capacity}  Due to the changing data distribution of user interests, an industrial-scale recommendation system's thirst for model capacity is insatiable. Especially in the early stage of training a recommendation model (\ie~the warm-up phase), larger model capacity is preferred in order to find a better local minimal and to reach the convergence desired.  As shown in Fig.~\ref{fig:pruning} (blue ellipse), a smaller model capacity (80\% sparsity) starts training from a window loss that is \textgreater0.225\% higher than the baseline.  
    \paragraph{Drifting data distribution} The input data distribution continuously shifts over time so that a fixed pruning mask during training  does not always adapt. As shown in Fig.~\ref{fig:pruning} (black ellipse), the training fails to converge when more and more data is fed to the pruned model. Thus, the sparsification should be reconstructed at intervals during training.
    
    \paragraph{Requirement on the final model} Recommendation models are usually trained offline before being launched to the recurring training with real-time data of user interests. It is required that the final model at the end of the offline training stage remains dense so that the recurring training stage possesses a large capacity to learn new emerging patterns from online data.

Hence, directly applying the pruning technique to recommendation models at scale causes significant performance degradation as compared to baseline, as shown in Fig.~\ref{fig:pruning}. 
Considering the constraints mentioned above, an ideal training scheme for recommendation systems at scale should start from a large capacity, leverages adaptive sparsification during training to reduce computation cost, and grows the model back to a full-size model for recurring training. 

In this paper, we propose a training scheme named alternate model growth and pruning. This scheme starts from training the full-size recommendation model, which benefits the warm-up stage. Once the full-size model converges,  important weights are selected through Taylor Approximation and then pruned.  Such a structured sparsity can largely reduce the computation cost. Once the sparse model converges, we grow the model to a full-size to avoid accuracy degradation caused by data distribution shifting. The growth and pruning occur by turns during training and conclude the offline training at a full-size model. Such a full-size model will be utilized in the recurring training and serving phase to achieve better performance. On the state-of-the-art deep learning recommendation model (DLRM) and a proprietary industrial-scale production models, we demonstrate that the proposed training scheme reduces 31.8\% and 15.3\% training FLOPs without hurting accuracy, respectively, without accuracy degradation.

To summarize, the contribution of this paper is three-folded:
\begin{itemize}
    \item We discuss and analyze the unique properties of modern recommendation models. These properties diminish the practicability and effectiveness of traditional model compression techniques such as pruning.
    \item To satisfy recommendation models' unique properties, we propose to alternately grow and prune the model during training, leveraging structured sparsification to reduce computation cost.
    \item The proposed scheme is validated with modern recommendation systems on the Criteo AI Labs Ad Kaggle dataset\footnote{https://www.kaggle.com/c/criteo-display-ad-challenge} and a real-world industrial dataset.

\end{itemize}
\section{Background}
% 2 column

\subsection{Recommendation Models}
%%% ================================================================================
\begin{figure}[!t]
    \centering
    \includegraphics[width=1.0\columnwidth]{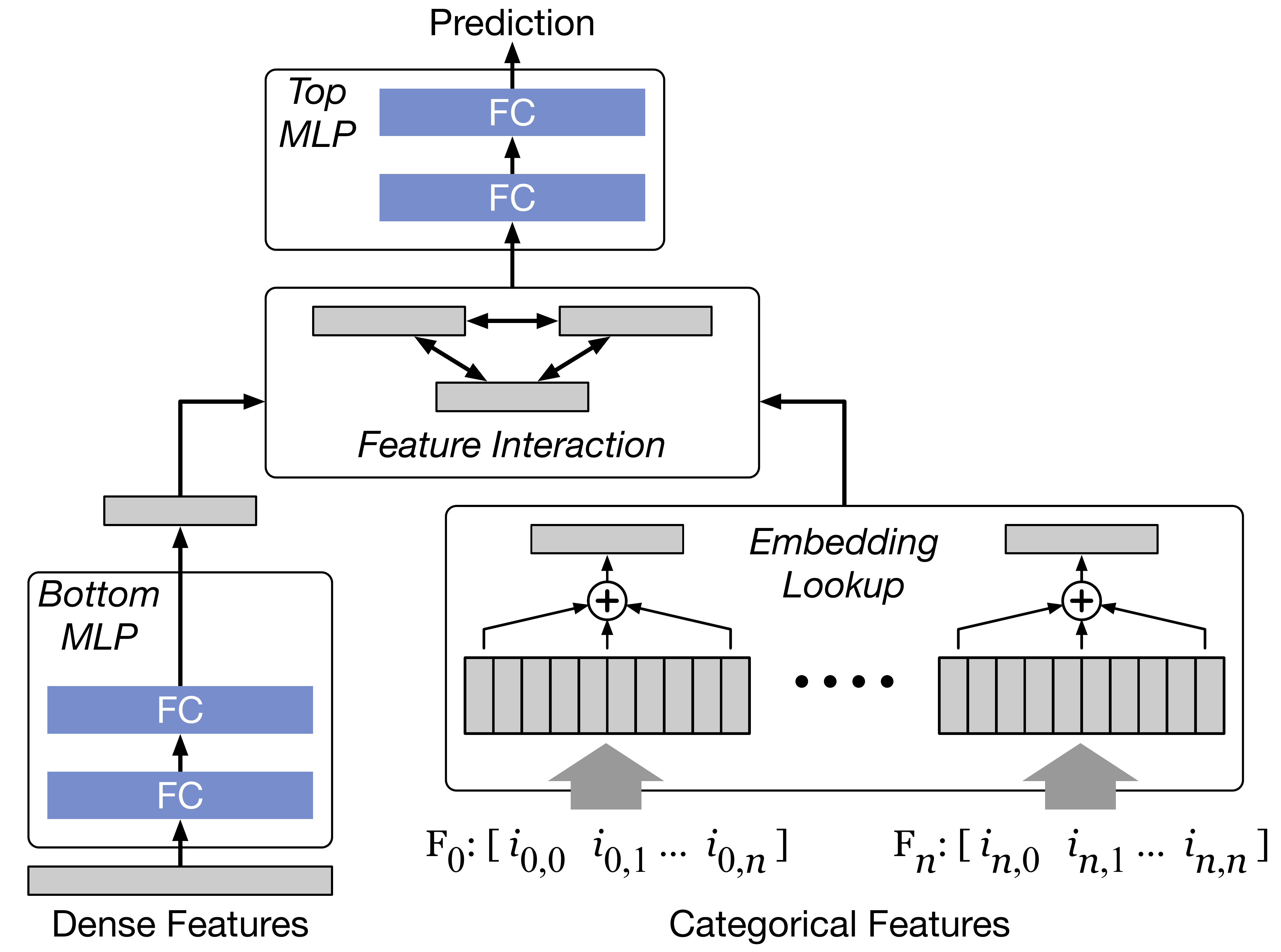}
    \caption{Structure of DLRM. Wide and deep fully-connected (FC) layers are employed.}
    \label{fig:dlrm}
\end{figure}
%%% ================================================================================

Modern recommendation models~\cite{naumov2019deep,cheng2016wide,zhou2019deep,du2019modeling} usually employ a structure shown in Fig.~\ref{fig:dlrm}. Categorical and continuous features are used as input and the most relevant content prediction for users is the targeted output. Categorical features are encoded into embedding lookup tables where each unique entity type (\eg~text/video/audio) is designated to a row of the embedding table, and multiple entities are pooled together and generate embedding tables via aggregate statistic. Continuous features are directly fed into a set of fully-connected layers (the bottom multi-layer perceptron). The categorical features and continuous features are coupled through a feature interaction, which is usually the dot product \cite{naumov2019deep}. Feature interaction is then fed into the top multi-layer perceptron (MLP), predicting the final output. In this paper, we demonstrate experiments on an open-source model named DLRM~\cite{naumov2019deep}, and a real-world industrial production model. Both models' structures  are similar to the structure mentioned above. Detailed setup can be found in Section~\ref{sec:exp_setup}.

%%% ================================================================================

\begin{figure*}[!t]
    \centering
    \includegraphics[width=0.9\textwidth]{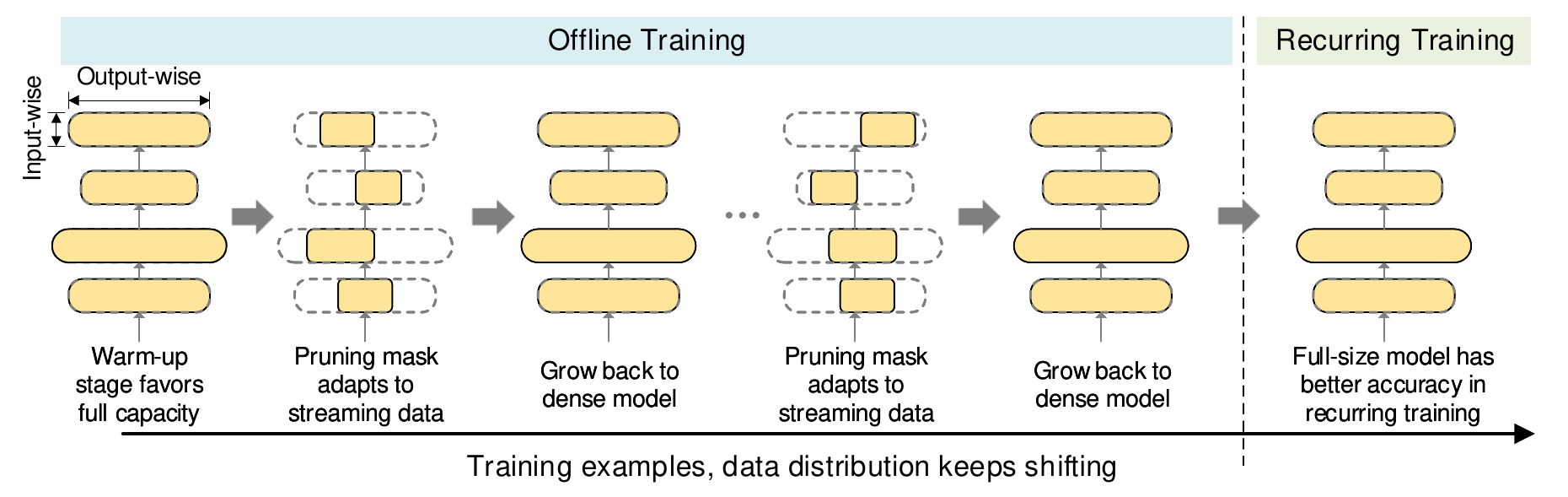}
    \caption{Overview of the alternate growth and pruning paradigm.}
    \label{fig:overview}
\end{figure*}
%%% ================================================================================

\subsection{Related Work} 
There have been broad interests in reducing the training cost of DNNs. In this section, we categorize and discuss them from the following three perspectives:

\paragraph{Network surgery}
Pruning during training has been extensively explored in computer vision area previously~\cite{he2014practical,han2015learning,liu2017learning,lebedev2016fast,alvarez2016learning,ye2020adaptive, kusupati2020soft,wang2019structured,hu2016network}. Pruning leverages network sparsity by removing less important parameters through either importance ranking~\cite{liu2017learning,han2015learning} or regularization~\cite{wang2019structured,hu2016network}. Despite the empirical success and the ability to learn the sparsity on general DNNs, pruning itself is not an optimal solution for recommendation systems due to the limitations discussed in Section~\ref{sec:intro}.

On the other side, constructive methods have also been explored in previous literature~\cite{ash1989dynamic,briedis1998using,dai2017nest,du2019cgap}. Constructive methods expand network structure or re-activate the pruned weights during training to incrementally increase model capacity, which is also referred to as model growth.
\cite{ash1989dynamic,briedis1998using} enlarged network capacity with fresh neurons and evaluated rudimentary problems such as XOR problems and multi-layer perceptron (MLP), without datasets from a real scenario. \cite{dai2017nest} picked a set of convolutional filters from a bundle of randomly generated ones to enlarge the convolutional layers. However,  finding one set of filters that reduces the most loss among several sets is through a trial-and-error way, consuming a significant amount of resources. Unlike previous methods, we simply embed alternate  growth and pruning of weights during training without extra efforts and apply it to real-world recommendation systems.

\paragraph{Incremental training}
Incremental training refers to splitting the entire training process into a couple of subsets or steps of training and then incrementally approaching the training goal. \cite{tao2019efficient} divides the learning concepts into different groups and trains the model with the sequential concepts. In  \cite{carta2020incremental},  the architecture is expanded by separating its hidden state into different modules, and new modules are sequentially added to the model to learn progressively longer dependencies. However, incremental training suffers from weight interfering and overwriting, \ie~new data overwrites the weights that are well trained by old data, especially when the data distribution drifts over time. Furthermore, incremental training usually requires extra memory and computation to reasonably divide training,  integrate sub-training, and retrieve previously learned weights.

\paragraph{Orthogonal methods}

There are several orthogonal methods, such as low-precision quantization and low-rank decomposition. The quantization technique compresses the DNN models by quantizing the parameters to fewer bits~\cite{gong2014compressing,hubara2017quantized,gupta2020fast}. The decomposition technique compresses the DNN models by finding a low-rank approximation~\cite{denton2014exploiting,leng2018extremely}. Note that the proposed scheme can be combined with these orthogonal methods to further improve training efficiency.

\section{Proposed Method}
In this section, we describe the proposed training flow. Recommendation models usually employ wide and deep \cite{cheng2016wide} fully-connected (FC) layers, which spend as much as over 40\% of total computation.
Thus, here we focus on improving the efficiency of FC layers. 
It is worth mentioning that, from the implementation point of view, the proposed training scheme is general and can also be applied to other types of architecture such as convolutional layers.

\subsection{Terminology}
A recommendation model can be treated as a feed-forward multi-layer architecture that maps the input data to certain output vectors. Each layer is a certain function, such as fully-connected, ReLU, and dot product. 
For the $l^{th}$ fully-connected layer, it is formulated as $\mathcal{Y}_l = \mathcal{X}_l \cdot \Theta_l$,    where the input $\mathcal{X}_l \in \mathbb{R}^{I_l}$, the output $\mathcal{Y}_l \in \mathbb{R}^{O_l}     \Leftrightarrow \mathcal{X}_{l+1} \in \mathbb{R}^{I_{l+1}}$, and the parameter matrix is $\Theta_l \in     \mathbb{R}^{O_l \times I_l}$. 

%%% ================================================================================

\begin{figure*}[!t]
    \centering
    \includegraphics[width=\textwidth]{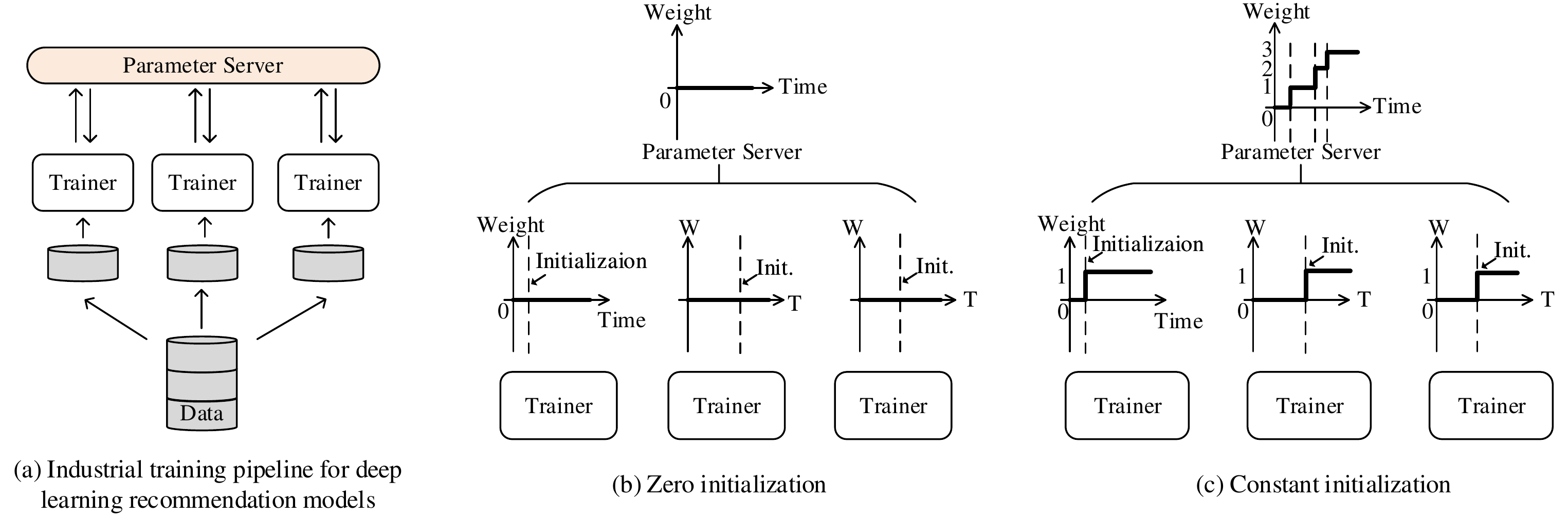}
    \caption{Zero initialization is the most robust given the industrial-scale training pipeline.}
    \label{fig:pipeline}
\end{figure*}
%%% ================================================================================

\subsection{Overview of the Proposed Scheme}

Fig.~\ref{fig:overview} presents an overview of the proposed training scheme. Unlike growth-only schemes~\cite{du2019cgap,yuan2020growing} that start the training from a small network seed, our method begins the training from a full-size model. Since the input data and feature interactions that need to be captured are in giant scales, a large model capacity for the warm-up stage is essential and can establish a better weight distribution and ultimately better model quality. After the initial model converges in the warm-up stage, we calculate an importance score (will be explained in the following subsection) for each neuron along the input-wise dimension,  then sort and preserve the most important  $100\times(1-\beta)\%$  neurons, where $\beta$ is the sparsity (\ie~the ratio of pruned weights compared to all the weights). Weights of the secondary neurons are set to zero, leading to a pruned model. This pruned model features a structured sparsity and largely reduced computation. When pruning takes place, a reasonable accuracy drop is usually observed.    Such an accuracy drop is gradually recovered along with the training on fine tuning data (\ie~the fine-tuning stage). However, if the pruning phase lasts too long, the sparse model starts to suffer from the insufficient capacity and the data drift problem as discussed in Section~\ref{sec:intro}, leading to worse accuracy. Therefore, our proposed method recovers the pruned model back to a full capacity (\ie~reactivate the pruned weights) in time to avoid the pruned model suffering from accuracy degradation. When the full-size model is converged, another iteration of importance evaluation and pruning is applied to reduce computation cost. These phases can be repeated until the end of offline training. In our proposed method offline training leads to a dense model that has sufficient capacity when deployed in the recurring training to learn the new emerging patterns in data in a real-time manner.

The proposed training scheme satisfies unique requirements of recommendation systems at scale, including (1) recommendation model is thirsty in capacity, especially for the warm-up stage; (2) pruning can be leveraged to reduce training cost but the pruning mask (the position of weights pruned) should not remain fixed throughout the training; (3) the training should complete at a full-size capacity to benefit recurring training.

\subsection{Model Pruning}
The pruning is performed in a structured way: all the input-wise weights connected to a neuron are concurrently pruned or preserved. Whether to keep or prune a specific neuron depends on its evaluation on the importance score, which is used to measure how important a neuron is to the loss function. Here, we use Taylor Approximation as the importance score, whose efficacy has been proven in \cite{score,du2020noise,ye2020adaptive}.
For a neuron $t$ in the $l^{th}$ layer $\Theta_l^t \in \mathbb{R}^{1\times I_l}$, the score is formulated as: 
\begin{align} \label{math:neuron_score}  
   |\Delta\mathcal{L}(\Theta_l^t)| \simeq |\frac{\partial{\mathcal{L}}(\mathcal{Y};\mathcal{X};\Theta)}{\partial{\Theta_{l}^t}}\Theta_{l}^t|   = \sum_{i=0}^{I_l}|\frac{\partial{\mathcal{L}}(\mathcal{Y};\mathcal{X}; \Theta)}{\partial{\Theta_l^{t, i}}} \Theta_l^{t, i}|, 
\end{align} 
where $\frac{\partial{\mathcal{L}}(\mathcal{Y};\mathcal{X}; \Theta)}{\partial{\Theta_l^{t, i}}}$ is the gradient of the loss with respect to the parameter $\Theta_l^{t, i}$.  
Based on the score above, we sort neurons layer by layer and identify the top $1-\beta$ neurons, where $\beta \in (0, 1)$. These relatively important neurons are preserved in the fine-tuning stage while the rest secondary neurons are pruned, generating $(\beta \times 100)\%$ structured sparsity.

\subsection{Model Growth}\label{sec:growth}

In the growth stage, the pruned weights are re-opened and allowed to learn, bringing the model back to full capacity. The pruned weights start updating from zero with the same learning rate. Zero initialization is effective yet straightforward for recommendation systems due to an asynchronous data-parallel distributed setup for industrial-scale recommendation systems. As shown in Fig.~\ref{fig:pipeline}a, large scale recommendation models are usually trained with multiple trainers (nodes) working on different partitions of data. Each trainer calculates gradients and synchronize to a centralized parameter server, while the server collects and averages the gradient and sends it back to each trainer for weights updating\cite{acun2020understanding}. For example, Hogwild\cite{recht2011hogwild} can be used to implement parallelizing stochastic gradient descent (SGD), leveraging multi-CPU/GPU to perform forward and backward pass. Zero initialization is able to get rid of asynchronous weight averaging sync problems caused by a distributed asynchronous SGD optimization setup, as shown in Fig.~\ref{fig:pipeline}b. In contrast, other types of initialization (such as constant initialization) could experience several sub-steps of unsmooth transition between zero and the initialization target, as shown in Fig.~\ref{fig:pipeline}c. 

From the implementation perspective, it is worth mentioning that when updating a huge number of parameters from zeros, denormal number\cite{denormal,denormal2} could be a severe problem. The mantissa in normal floating point is normalized and does not have leading zeros. However, there are leading zeros in the mantissa in denormal numbers, which would result in an exponent that is too small and not representable.  Denormal numbers can incur extra computational cost and significantly slow down the computation speed\cite{stackoverflow}. Setting both the flush-to-zero (FTZ) and denormals-are-zero (DAZ) flags to true could solve this problem\cite{intel}.

\section{Experimental Results}
We use both an open-source DLRM\footnote{https://github.com/facebookresearch/dlrm} model and a  real-world industrial recommendation model to perform experiments. 

\subsection{Experiment Setup}\label{sec:exp_setup}

\paragraph{Training Setup}

For the DLRM model, which follows the structure discussed in Fig.~\ref{fig:dlrm}, the embedding dimension for sparse features is set as 4. The output dimensions of the bottom FC-layer stack are 256, 128, 64 and 4, respectively, while the top FC-layer stack are 256, 128 and 1. Experiments are performed on a  single Nvidia Quadro RTX 8000 platform with PyTorch.  Stochastic gradient descent (SGD) is adopted with a constant learning rate of 0.1.
For the industrial recommendation model, the structure is similar to the architectures mentioned above, and the training is performed on parallel trainers as discussed in Section~\ref{sec:growth}. Adagrad~\cite{duchi2011adaptive} is used as the optimizer.

\paragraph{Dataset}
Few public data sets are available for recommendation and personalization systems. On DLRM, we use  Criteo AI Labs Ad Kaggle dataset\footnote{https://labs.criteo.com/2014/02/kaggle-display-advertisingchallenge-dataset/} to perform experiments. This dataset contains 13 sparse features and 26 categorical features, approximately 45 million samples over 7 days. In our experiments, 6 days are used as the training set and the rest 1 day as the testing set.
On the industrial-scale recommendation model, we use a proprietary real-world dataset. Both datasets are trained with a single pass, \ie~one epoch.

\paragraph{Evaluation}
For the DLRM model, binary cross-entropy (BCE) is used as the loss function. Both training and testing accuracy are reported. When comparing the proposed method with the baseline model, we also report the relative accuracy:
\begin{equation}
 \resizebox{0.9\columnwidth}{!}{$
    \text{relative } \mathrm{accuracy} =\frac{\text {Accuracy of the experimental model}}{\text {Accuracy of the baseline model}}-1
$}
\end{equation}
The industrial model uses the normalized cross-entropy (CE) \cite{he2014practical} metric from the classification task, which measures the average log loss divided by what the average log loss would be if a model predicted the background click-through rate. Since for recommendation system, we care more about the performance on the incoming data rather than the entire historical data. We report a performance metrics named look ahead window evaluation (eval) CE \cite{cervantes2018evaluating}, where CE is measured with the unseen incoming data within a given moving time window.
In the following experiments, we define the relative CE loss as below:
\begin{equation}
 \resizebox{0.75\columnwidth}{!}{$
    \text{relative } \mathrm{CE} =\frac{\text{CE of the experimental model}}{\text {CE of the baseline model}}-1 
$}
\end{equation} 
 
Usually,  $|\text{relative CE/accuracy}|$\textgreater 0.1\% is considered as a significant difference.
We use training FLOPs to estimate the training computational cost for DLRM and industrial model.

\subsection{Initial Model Capacity}
\begin{figure}[!t]
    \centering
    \includegraphics[width=0.85\columnwidth]{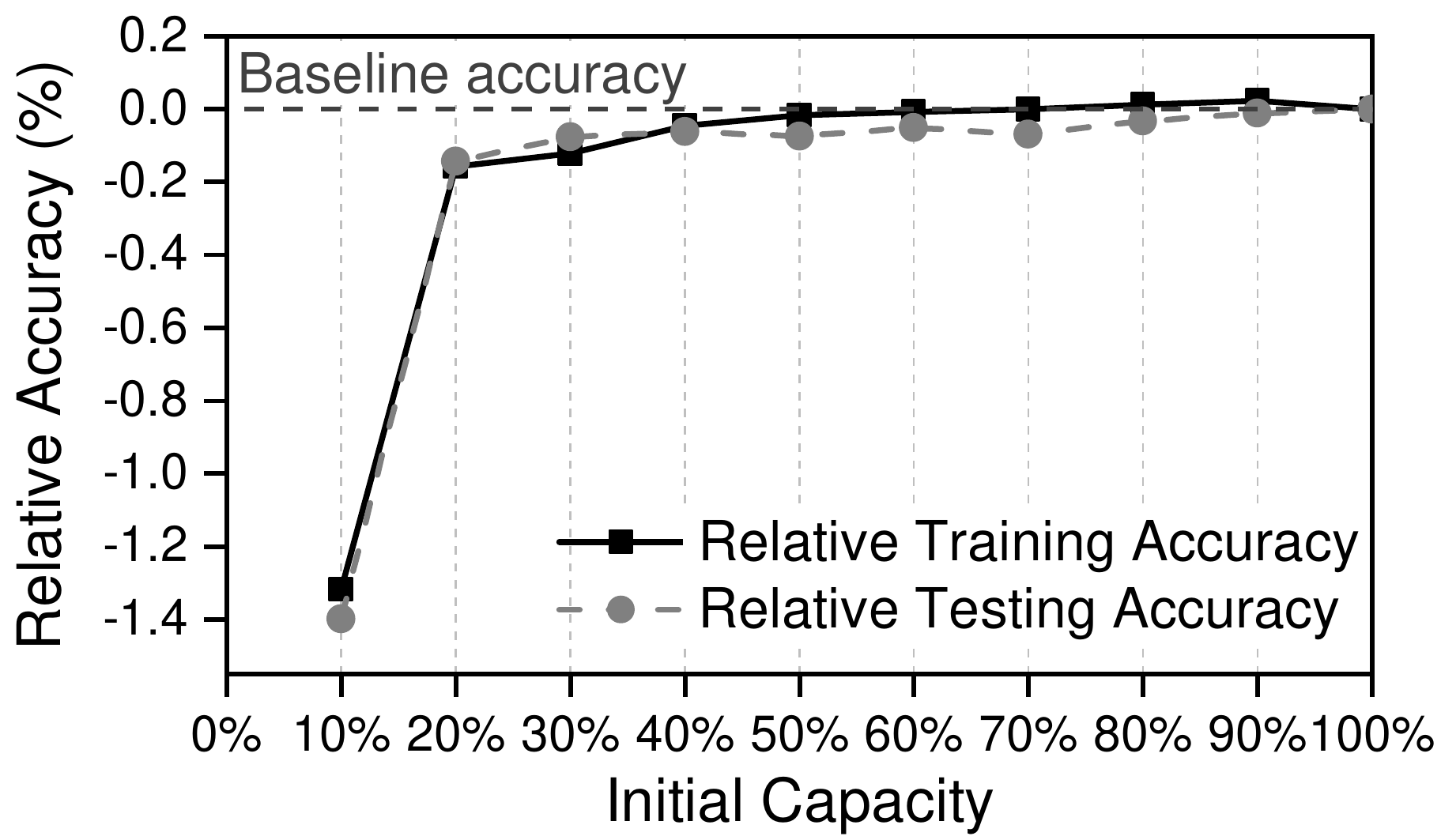}
    \caption{Relative accuracy (\%) for models with different initial capacity.}
    \label{fig:initcap}
\end{figure}

As discussed above, a larger model capacity is preferred at the early stage of training. To demonstrate how the warm-up phase influences the final accuracy, we start the training from a smaller model capacity (\eg~10\%, 20\%, ..., 90\% as compared to the full size) for the first 10\% samples, and then grow the models back to full size and train with the rest of the samples. 
In Fig.~\ref{fig:initcap}, we plot the relative accuracy for various initial capacities. With larger initial capacity, the final training and testing accuracy both improve. This observation validates that a dynamic training scheme should start from a dense model rather than a sparse model, though sparse-to-dense scheme can reach the same computation saving as compared to the proposed scheme.

%%% ================================================================================
%%% ================================================================================

\subsection{Validating the Proposed Scheme}\label{sec:valid}

\begin{figure*}[!t]
\begin{center}
\includegraphics[width=0.9\textwidth]{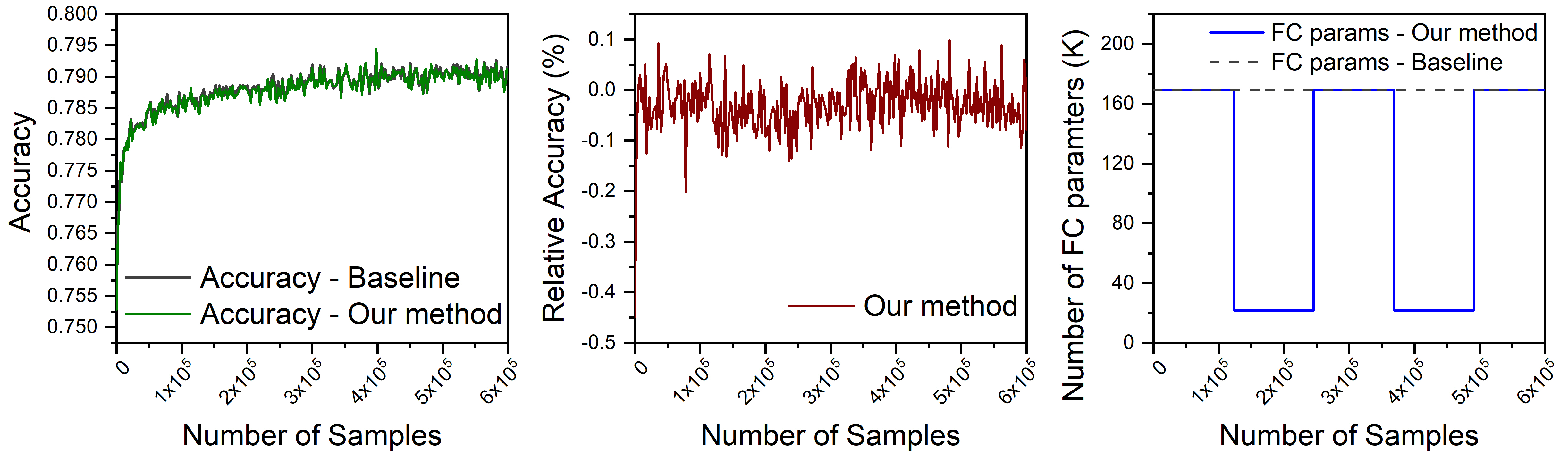}
\caption{Training accuracy, relative accuracy and the number of total parameters in FC layers during training. Our scheme is on par with the baseline model in accuracy and loss, with additional saving in training FLOPs and number of parameters. }
\label{fig:learning_curve}
\end{center}
\end{figure*}

In Table~\ref{tab:acc}, we present the training accuracy, testing accuracy, training FLOPs  of the proposed scheme under different sparsity $\beta$, and their relative change as compared to the baseline. In the experiments, each growth and pruning phase is trained alternatively with 20\% of the training samples, under the condition that the training starts from a dense model, alternates between growth and pruning during training, and ends at a dense model. On the DLRM model, the alternative growth and pruning scheme is able to achieve neutral accuracy with up to $\sim$32\% FLOPs saving in a sparsity of 70\%, as compared to baseline. In Fig.~\ref{fig:learning_curve}, we present the training accuracy, the relative training accuracy as compared to baseline, and the number of total FC parameters during training.

%%% ================================================================================
\begin{table*}[!t]
\caption{Comparing the alternative growth and pruning scheme to the baseline on the DLRM model. Note that a $|\text{relative accuracy}| > 0.1\%$ is considered as a significant change.}
\label{tab:acc}
\centering
  \resizebox{0.9\textwidth}{!}{  
\begin{tabular}{c|cc|cc|cc}
\hline
\textbf{Sparsity $(\beta\times100)$ } & \textbf{Training Accuracy} & \textbf{Relative} & \textbf{Testing Accuracy} & \textbf{Relative} & \textbf{Training FLOPs ($\times10^6$)} & \textbf{Relative} \\ \hline
\textbf{Baseline} & 0.7901                     & -                 & 0.7866                    & -                 & 207.7                                                                & -                 \\ \hline
90\%              & 0.7900                     & -0.02\%           & 0.7859                    & -0.08\%           & 129.5                                                                & -37.7\%           \\
80\%              & 0.7902                     & 0.02\%   & 0.7859                    & -0.09\%  & 135.3                                                                & -34.9\%  \\
70\%              & 0.7906                     & \textbf{0.06\%}            & 0.7864                    & \textbf{-0.02\%}           & 141.6                                                                & -31.8\%           \\
60\%              & 0.7904                     & 0.03\%            & 0.7864                    & -0.03\%           & 148.8                                                                & -28.4\%           \\
50\%              & 0.7905                     & 0.04\%            & 0.7863                    & -0.04\%           & 156.6                                                                & -24.6\%           \\
40\%              & 0.7904                     & 0.04\%            & 0.7860                    & -0.08\%           & 162.8                                                                & -21.6\%           \\
30\%              & 0.7905                     & 0.05\%            & 0.7861                    & -0.07\%           & 164.4                                                                & -20.8\%           \\
20\%              & 0.7904                     & 0.04\%            & 0.7863                    & -0.03\%           & 164.9                                                                & -20.6\%           \\
10\%              & 0.7905                     & 0.05\%            & 0.7863                    & -0.04\%           & 165.5                                                                & -20.3\%           \\ \hline
\end{tabular}
}
\end{table*}
%%% ================================================================================
% figure: learning curve/FLOPs: DSDSD vs baseline
% table: different sparsity, accuracy/flop, dsdsd vs baseline

%%% ================================================================================
%%% ================================================================================

%%% ================================================================================

\begin{table}[!t]
\caption{Comparison among 4 typical training schemes on DLRM.}
\label{tab:4sheme}
\centering
  \resizebox{0.95\columnwidth}{!}{  
\begin{tabular}{c|cc|c}
\hline
\textbf{Scheme}                         & \textbf{Accuracy} & \textbf{Relative} & \multicolumn{1}{l}{\textbf{\begin{tabular}[c]{@{}l@{}}Training FLOPs\\~~~~~~($\times10^6$)\end{tabular}}} \\ \hline
\textbf{Baseline}                       & 0.7866            & -                 & 207.7                                       \\ \hline
\textbf{Pruning-only}                   & 0.7861            & -0.06\%           & \multirow{3}{*}{}                       \\ \cline{1-3}
\textbf{Growth-only}                    & 0.7858            & -0.10\%            &   141.6                                 \\ \cline{1-3}
\textbf{DSD~\cite{han2016dsd}}             & 0.7858            & -0.10\%    &        (-34.9\%)                               \\\cline{1-3}
\textbf{Our method} &\textbf{0.7864}            & \textbf{ -0.02\%}           &                                       \\\hline
\end{tabular}
}
\end{table}
%%% ================================================================================

\subsection{Comparison with Other Schemes}
We design experiments to compare the alternative growth and pruning method with some other commonly used training scheme, including pruning-only, growth-only, dense-sparse-dense (DSD)~\cite{han2016dsd}. The experiments are designed as below: for each scheme, we use $\beta=70\%$ as the sparsity, which is the upper bound of sparsity to achieve neutral accuracy for the proposed method.  The sparsification lasts for 40\% training samples for all the schemes. In other words, for the pruning scheme, the first 60\% training samples are trained with a full-size model while the rest 40\% training samples are trained with a sparse model, and the pruning is based on Taylor Approximation; Similarly, for the growth scheme, the first 40\% training samples are trained with  a sparse model while the rest samples are trained with a full-size model, and the growth starts with zero initialization; for the DSD scheme, the first 30\% and the last 30\% samples are trained with a dense model, while the middle 40\% samples are trained with a sparse model; for our alternative growth and pruning scheme, the setup is the same as discussed in Section~\ref{sec:valid} and Fig.~\ref{fig:learning_curve}. All the other experimental setups are the same across four schemes to guarantee a fair comparison.

We present the results of four schemes on DLRM in Table~\ref{tab:4sheme}.  With the same  training FLOPs, our method outperforms the other three schemes in accuracy. The reason is that the alternative growth and pruning scheme (1) starts training from full capacity, benefiting the warm-up stage of the recommendation model; (2) each sparse phase during training does not last too long to avoid performance degradation; (3) the model ends at full capacity.

%%% ================================================================================
\begin{figure*}[!t]
\begin{center}
\includegraphics[width=0.9\textwidth]{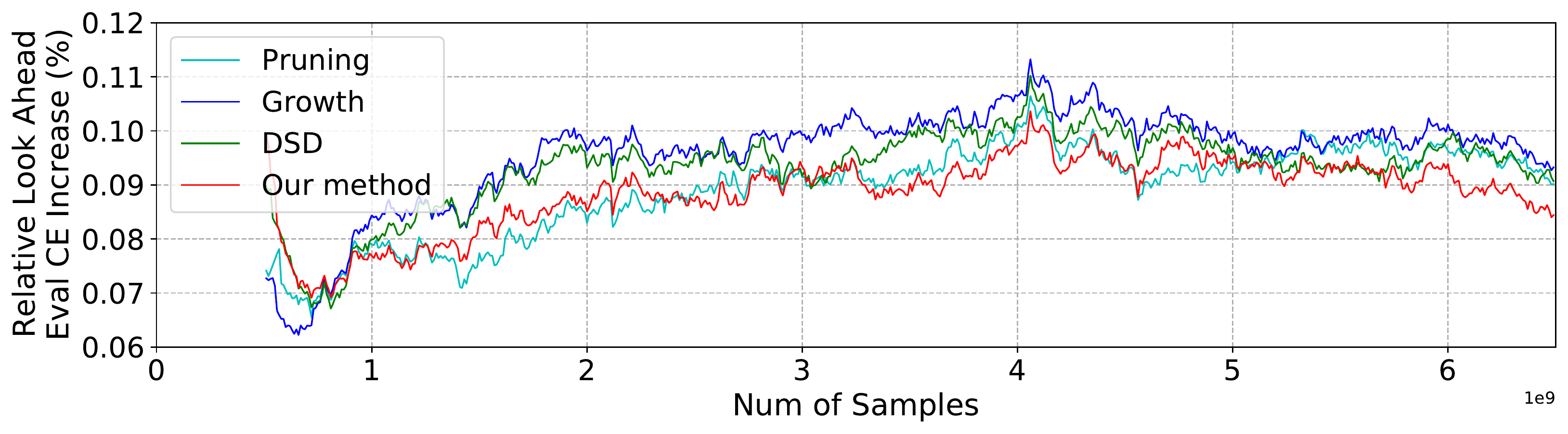}
\caption{Comparing CE loss among different training schemes on industrial model. }
\label{fig:indust}
\end{center}
\end{figure*}
%%% ================================================================================

Similarly, we provide the relative look ahead evaluate CE loss for four schemes on the industrial model, as shown in Fig.~\ref{fig:indust}. All the schemes are set with $\beta=30\%$. The pruning scheme is trained with full capacity with the first 70\% training samples and then pruned; the growth-only scheme is trained with a small capacity on the first 30\% samples and then recovered back to full size; for the DSD scheme, the middle 30\% training samples (\ie~35\% to 65\%) are trained with sparse model; for our method, the 20\% to 35\% and 50\% to 65\% (overall 30\%) training samples are trained with sparse model. With this setting, all the schemes have 15.3\% less training FLOPs. From Fig.~\ref{fig:indust}, it is concluded that our method achieves the lowest loss (0.084\% as compared to the baseline), outperforming the other three schemes in a industry-scale recommendation model.

\section{Conclusion}

In this work, we provide comprehensive understanding on unique properties and requirements of leveraging sparsification to save computation cost for recommendation models at scale. Correspondingly, we propose the alternative growth and pruning scheme to reduce FC computation during training. On DLRM model and a real-world industrial model, we demonstrate that the proposed scheme is able to reduce 31.8\%  and 15.3\% training FLOPs without performance degradation. The proposed scheme outperforms both pruning-only schemes and growth-only schemes since it considers the unique properties of large-scale recommendation models. In the future, we plan to augment the proposed scheme with more advanced model surgery techniques and/or initialization techniques to further improve its performance.

% \bibitem{b1} G. Eason, B. Noble, and I. N. Sneddon, ``On certain integrals of Lipschitz-Hankel type involving products of Bessel functions,'' Phil. Trans. Roy. Soc. London, vol. A247, pp. 529--551, April 1955.
% \bibitem{b2} J. Clerk Maxwell, A Treatise on Electricity and Magnetism, 3rd ed., vol. 2. Oxford: Clarendon, 1892, pp.68--73.
% \bibitem{b3} I. S. Jacobs and C. P. Bean, ``Fine particles, thin films and exchange anisotropy,'' in Magnetism, vol. III, G. T. Rado and H. Suhl, Eds. New York: Academic, 1963, pp. 271--350.
% \bibitem{b4} K. Elissa, ``Title of paper if known,'' unpublished.
% \bibitem{b5} R. Nicole, ``Title of paper with only first word capitalized,'' J. Name Stand. Abbrev., in press.
% \bibitem{b6} Y. Yorozu, M. Hirano, K. Oka, and Y. Tagawa, ``Electron spectroscopy studies on magneto-optical media and plastic substrate interface,'' IEEE Transl. J. Magn. Japan, vol. 2, pp. 740--741, August 1987 [Digests 9th Annual Conf. Magnetics Japan, p. 301, 1982].
% \bibitem{b7} M. Young, The Technical Writer's Handbook. Mill Valley, CA: University Science, 1989.
% \end{thebibliography}
% \vspace{12pt}
% \color{red}
% IEEE conference templates contain guidance text for composing and formatting conference papers. Please ensure that all template text is removed from your conference paper prior to submission to the conference. Failure to remove the template text from your paper may result in your paper not being published.
% % \section{Reference}
\bibliographystyle{IEEEtran.bst}
\bibliography{ref}

\end{document}